\documentclass{optica-article}

\journal{opticajournal} 

\articletype{Research Article}

\def\ttitle{Iterative Refinement of Arbitrary Micro-Optical Surfaces}

\usepackage{lineno}
\usepackage{hypernat}
\usepackage[nolist]{acronym}

\usepackage{array}
\usepackage{amsmath}
\usepackage{./styles/shortcuts} 
\usepackage{eso-pic}
\usepackage{
color}
\usepackage{type1cm}
\usepackage{import}      
\usepackage{mathtools, nccmath}
\usepackage{subfigure}
\usepackage{pgf}

\begin{acronym}
  \acro{ROC}{radius of curvature}
  \acro{FFPC}{fiber Fabry-Perot cavity}
  \acro{FWHM}{full width half maximum}
  \acro{RMS}{root-mean-square}
  \acro{cQED}{cavity quantum electrodynamics}
  \acro{PSI}{phase shifting interferometry}
  \acro{SM}{single mode}
  \acro{MM}{multi-mode}
  \acro{GRIN}{gradient index}
  \acro{PC}{photonic crystal}
  \acro{AOM}{acousto-optic modulator}
  \acro{PD}{photodiode}
\end{acronym}

\begin{document}

\title{\textbf{\ttitle}}

\author{Meagan Plummer,\authormark{1,3} Stephen Taylor,\authormark{4} Matthew Marshall,\authormark{2,3} David Brown,\authormark{4} Robert Leonard,\authormark{5} Seth Hyra,\authormark{6} and Spencer Olson\authormark{6,*}}

\address{\authormark{1}Optical Science \& Engineering Program, University of New Mexico, Albuquerque, NM 87131\\
\authormark{2}Department of Physics and Astronomy, University of New Mexico, Albuquerque, NM 87131\\
\authormark{3}Universities Space Research Association\\
\authormark{4}National Research Council, National Academies of the Sciences\\
\authormark{5} Space Dynamics Laboratory, Quantum Sensing \& Timing, North Logan, UT 84341, USA\\
\authormark{6}Air Force Research Laboratory, Kirtland Air Force Base, NM 87117, USA}

\email{\authormark{*}spencer.olson.2@afrl.af.mil} 

\date{}

\newcommand{\distA}[1]{%
  Approved for public release; distribution is unlimited.  Public Affairs %
  release approval %
  #1.
}

\pagestyle{fancy}
\fancyhead{}
\renewcommand{\headrulewidth}{0pt}
\fancyfoot{}
\fancyfoot[R]{\thepage}
\fancyfoot[L]{
  \footnotesize
  \centering{\distA{\#AFRL-2024-7002}}
}

\begin{abstract} 

We introduce an adaptive optical
refinement method enabling ultra-precise micro-milling of arbitrary surfaces.
Through repeated iteration, our method reduces surface error without requiring significant specific
surface engineering. This remediates the long sample preparation times and lack of refinement
capability that previously reported methods suffer from. The iterative refinement milling method was used to produce spherical mirrors with
small radii of curvature and low surface roughness for use in micro Fabry-Perot
cavities.  We demonstrate the use of this adaptive process to produce a
variety of arbitrary surface geometries on both optical fiber tips as well as optical
flats.  We additionally discuss our capability to apply iterative refinement milling adaptively
to various materials, including to construct GRIN lenses.
\end{abstract}

\clearpage
\section{Introduction}

The development of advanced micro-milling techniques has contributed to the ubiquity of high finesse optical cavities, including \acp{FFPC}, in applications across
many disciplines, from sensing to information~\cite{Miller1990, Toninelli2010,
Mader2015, Ma2023, Cooper2024}.
\acp{FFPC} are micro-optical resonators made by shaping the tips
of optical fibers such that stable cavities can be formed in the gap between
two opposing tips~\cite{Pfeifer2021}. \acp{FFPC} are well-suited for strong
atom-cavity coupling due
to their small mode-volume, inherent fiber coupling, and open-access
geometry. However, traditional polishing techniques are insufficient
to form mirrors with the small radii of curvature required to form a stable
\ac{FFPC}~\cite{Pfeifer2021}. Several techniques have been developed to machine
concave surfaces on optical fibers, including chemical etching~\cite{Qing2019},
focused ion beam milling~\cite{Trichet2015}, and laser ablation~\cite{hunger_fiber_2010}.
Of these, laser ablation has been the most promising to produce sufficiently low
roughness surfaces to enable high finesse cavities~\cite{Pfeifer2021}.

Laser ablation milling is typically performed by applying a single pulse from a
$\rm CO_2$ laser to the facet of an optical fiber, creating a Gaussian-like
indentation on the fiber~\cite{hunger_fiber_2010}. A drawback to this technique is that the size and
shape of the indentation is limited by the transverse mode of the ablating
beam~\cite{Ruelle2019}. This constraint can be removed by applying a series of
weak pulses to spatially separate points, or dots, across the fiber surface~\cite{Ott2016}.
This technique, sometimes referred to as \textit{dot-milling}, may be used to mill
near-arbitrary surface geometries.

A significant drawback of the dot-milling technique is that the power, duration, and location 
of each spatially separate pulse must be tuned precisely to minimize error in
the milled surface~\cite{Ott2016}. The parameter space which must be explored to
optimize this milling process is extensive. These parameters are
highly coupled, as the indentation formed by each pulse overlaps with indentations of
neighboring pulses.  Furthermore, the efficacy of a
laser pulse to cause material removal is significantly affected by small
deviations in the local surface shape where the laser interacts.  Additionally, changes in
environmental conditions can alter the laser-surface interaction and affect material removal.
Moreover, learning by iterating through parameters applied to a known starting surface
requires samples to be (re)polished and/or replaced frequently.  Finally, such an
optimized process will need to be re-engineered whenever the target surface geometry or
the target material changes.

In this paper we present an adaptive milling method where
multiple iterations hone the surface shape and asymptotically correct for surface
errors.  The primary effect of this approach is that specific optimization
of the temporal and spatial milling pattern is not necessary.  Furthermore, this
approach can be easily applied to varying materials and surface geometries,
provided a few engineering restrictions are met.
Effectively, our process can be described by a feedback loop iteratively
integrated with a dot-milling technique, superficially de-coupling the
complexities of material physics and laser milling to achieve an optimized,
error reduced, target surface.

\section{Surface Quality as a Figure of Merit}\label{sec:roughness}

As the techniques presented in this paper allow for a variety of functional
forms for the target surface, the actual performance of the milled
surface will ultimately depend on the specific application.  Nevertheless, as a
technique to apply generically to optical surfaces and for the scope of this
paper, we simply focus on statistics of the measured difference of the
milled surface to the desired functional form. This includes both full surface, macroscopic geometric deviations, as well as surface roughness. 

As an example of this simple metric for evaluating milled-surface quality,
consider the surfaces used to create \acp{FFPC}.
A clear defining measure of quality for
\acp{FFPC} is their finesse, an indicator of the sharpness of the etalon
interference fringes and directly proportional to the single-atom
cooperativity~\cite{Colombe:2007:saf}, as related to \ac{cQED}.  While often
measured in terms of free spectral range divided by the \ac{FWHM} of the
resonant peaks, the finesse of a cavity, $\mathcal{F}$, can be represented
solely as dependent on losses, $\mathcal{L}$, in the cavity
via~\cite{Ismail:2016}
  
\begin{equation}
    \mathcal{F}(\mathcal{L}) = \frac{\pi}
                                    {2\sin^{-1}\left(
                                      \frac{1-\sqrt{1-\mathcal{L}}}
                                           {2\sqrt[4]{1-\mathcal{L}}}
                                    \right)}.
\label{eq:finesse}
\end{equation}

\begin{figure}[ht]
    \centering
    \input{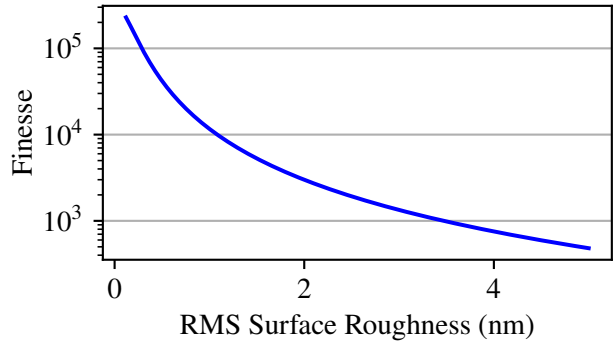}
    \caption{Scattering limited finesse for \ac{RMS} surface roughness from 0 to 5~\nm for a nominal confocal cavity design with 1~\mm \ac{ROC}.}
    \label{fig:finesse-roughness}
\end{figure}

One can then determine the highest potential finesse of a
system in terms of its most limiting loss source, assuming that the worst loss
source is much more significant than others. The primary
loss mechanisms in \acp{FFPC} are mode clipping,
absorption, transmission, and surface scattering. For many geometries, it is
relatively straight forward to minimize clipping by
increasing mirror size, and in the case where the cavity
mirror diameters are $\geq$ 3$\times$ the
\ac{FWHM} of the transverse cavity mode at the mirror
surface, clipping losses become negligible (on the order
of $0.004\,{\rm ppm}$).  Transmission and
absorption losses are dictated by the quality of the
reflective coating applied to the cavity mirrors,
something outside the purview of the present discussion. 
Additional impacts of coating  and/or annealing on final surface roughness are also out of scope
of the present paper. As such, the best figure of merit
connecting fabrication quality to cavity finesse is the optical
surface roughness. The \ac{RMS} surface roughness value,
$\sigma$, can be directly connected to a scattering loss
for a given wavelength, $\lambda$, by \cite{Bennett:1978}
\begin{equation}
    \mathcal{L}_s = 1-e^{-\left( \frac{4 \pi \sigma}{\lambda} \right)^2} \approx \left( \frac{4 \pi \sigma}{\lambda} \right)^2,
\label{eq:scatter}
\end{equation}
and therefore to a scattering limited finesse $\mathcal{F}(\mathcal{L}_s)$, see~Fig.\ref{fig:finesse-roughness}.

From Eqs.~\ref{eq:finesse} and~\ref{eq:scatter}, order of magnitude relations
can be made between surface roughness and scattering limited finesse. For
instance, taking a confocal cavity configuration with millimeter \ac{ROC}, a
surface roughness $\sigma = 1 \nm$ leads to $\mathcal{L}\approx520\,{\rm ppm}$ and
$\mathcal{F}\approx11,000$, whilst a $\sigma = 0.5 \nm$ leads to
$\mathcal{L}\approx130\,{\rm ppm}$ and $\mathcal{F}\approx40,000$. Within the context of
\ac{cQED}, the strong coupling regime occurs for single atom cooperativity $>1$, where cooperativity is given by~\cite{Motsch2010}
\begin{equation}
    \eta = \frac{6}{\pi^3} \frac{\mathcal{F} \lambda^2}{w_0^2}
\label{eq:atomcoop}
\end{equation}
In this example, the strong coupling regime is achieved when $\mathcal{F}\succsim1,000$, 
and therefore when surface roughness $\sigma \leq 3.38$~\nm. Different
applications will imply different finesse and atom cooperativity requirements, similarly setting bounds
for maximum surface roughness. Cooperativities in
excess of $100$, deep within the strong coupling regime, correspond in our example system to a confocal millimeter cavity finesse
of $\mathcal{F} = 110,000$ and a surface roughness  $\sigma \leq 0.27 \nm$. 

To note, atom cooperativity may also be improved
by reducing mode volume (proportional to the mode radius, $w_0$), as can be
achieved through approaching the concentric regime. This will not directly
impact the finesse, though clipping loss considerations will drive a requirement for
larger radii mirrors to maintain a finesse commensurate with a desired atom
cooperativity. The near concentric regime can therefore allow for higher atom
cooperativities with a lower finesse, and thus a relaxed
constraint on surface roughness. 
Surface roughness remains the limiting manufacturing parameter for both finesse and atom cooperativity, however, and as such is our figure of merit when assessing mill quality. 

Minimization of macroscopic \ac{RMS} deviations from a desired target surface also represent  mill quality, indicating the degree of control over final milled surface geometry. The impacts of these surface errors are more specific to target geometry and application, and a thorough mode analysis for all milled structures is beyond the scope of this paper. Values for overall surface \ac{RMS} will be provided as additional figures of merit for various milled geometries, and compared with those achieved by other fabrication methods.

\section{Iterative Refinement Milling}\label{sec:iterative-refinement}

\begin{figure}
    \centering
        \input{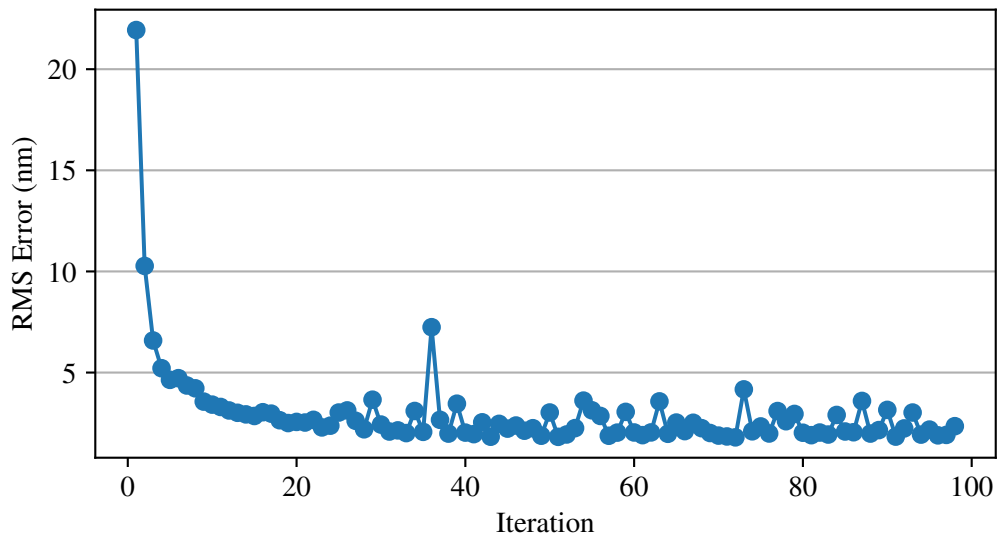}
    \caption{
        Fixed \ac{RMS} error rapidly declines during the first few iterations of a 1\mm \ac{ROC} spherical surface.
        As the milled surface converges towards the targeted surface, the fixed
        \ac{RMS} error continues a slow and steady decline.  After approximately
        25 iterations (dependent on the geometry of the target surface), we reach the limits of the milling precision, and
        \ac{RMS} error begins to vary about a steady-state value.
        Providing an exit condition for our mill code enables consistent achievement of low \ac{RMS} results. This example is characteristic of milling behavior.  
     }
    \label{fig:rms_it.}
\end{figure}

Iterative refinement milling involves successive milling operations over a given
surface to minimize overall error between the current measured surface and the final target surface.
This error,
measured using an optical
profilometer, is input into a proportional feedback loop,
where a set of laser pulse parameters are
calculated to minimize the error.
These corrective laser pulses are applied using a tightly
integrated $\rm CO_2$ laser system.  The resulting surface feeds back
into the refinement loop, where the new error is measured.  This
process is repeated, resulting in asymptotically reduced surface
error, see Fig.~\ref{fig:rms_it.}.

\begin{figure}[h]
    \centering
    \includegraphics[width=\columnwidth]{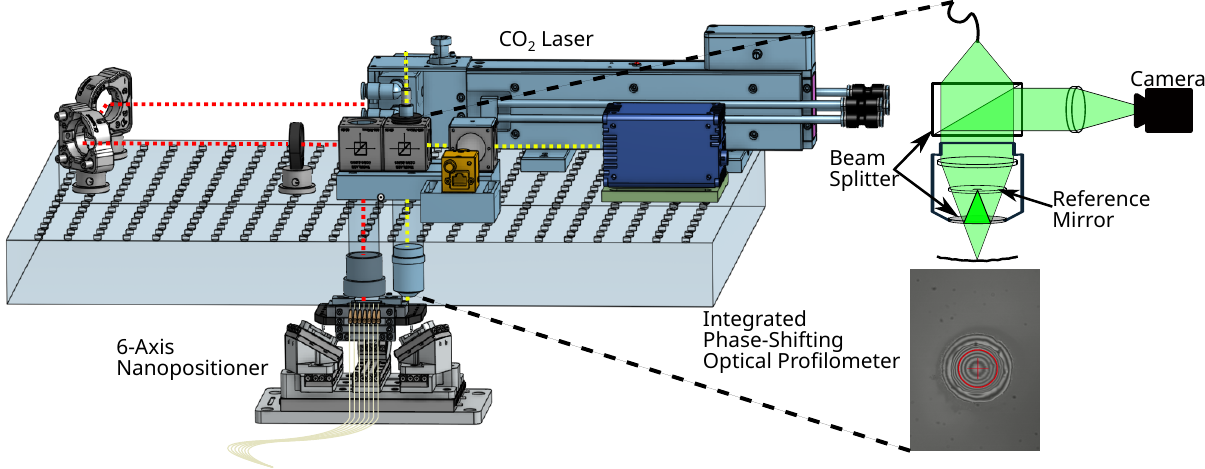}
    \caption{
        Schematic of the laser milling system, showing the tight integration
        between the beam path for the laser ablation and the optical profilometer.
    }
    \label{fig:optics_schematic}
\end{figure}

\subsection{Theory of Operation}\label{sec:theory-of-ops}

The target surface is established by a user-defined function which is called by the	refinement code. These surfaces are mapped to a hexagonal grid of equidistant points which set locations for individual mill pulses. The choice of grid spacing between points, $\delta r$, impacts the mill quality and speed. We additionally shift the grid by a random fraction of $\delta r$ between each mill iteration to prevent accumulation of grid structured systematic error. We have demonstrated milling spherical surfaces, multiple closely positioned spherical surfaces, and conical axicon surfaces, as discussed in Sec.~\ref{sec:results}.

We have additionally developed a simulated 
version of our custom laser mill code. This enables rapid trial and 
implementation of code improvements and new target surface geometries 
before deployment to the hardware. Trial arbitrary surfaces can be tested on the simulation to identify any required major alterations to milling parameters or surface definition before attempting destructive testing. This additionally allows for distinction between physical and algorithmic issues in milling.

Our system possess insufficient delivered power to produce clean ablation, and is 
observed to re-deposit material near the mill site. 
Our iterative method removes the necessity for clear understanding of the 
surface light-matter interactions, however, and can accommodate for these effects. Iterative refinement milling is adaptable to various power regimes of ablative laser milling. 

The function of our iterative refinement technique relies upon the ability to reduce errors in proportion with the amount of material removed during milling. Sources of milling error can therefore be appropriately sorted into two categories:
\textit{constant errors} and \textit{proportional errors}, where 
proportional errors are strongly correlated with the volume of 
material removed. Proportional errors dominate milling error during 
the initial iterations. During each iteration, a set of laser pulses removes a volume of 
material which is proportional to the difference between the target 
surface and the current surface.  Consequently, less volume is removed 
during subsequent iterations, thereby reducing the proportional 
errors.  This process continues until the milling error arising from 
proportional error sources are commensurate with the milling error 
from constant error sources.

The milling apparatus, see Fig.~\ref{fig:optics_schematic} has been designed to minimize both constant and proportional errors. In the following sections, we will address each source of error separately, as well as feedback stability considerations.

\subsection{Proportional Errors}

Proportional errors limit the rate of convergence while maintaining 
feedback stability (we will discuss stability in 
Sec.~\ref{sec:stability}).  Indeed, in the absence of proportional 
errors, we could achieve minimum error through a single dot-milling 
iteration. Proportional errors include calibration error, pulse duration noise, and laser power noise. 

\subsubsection{Calibration Error}
\label{sec:calibration}

\begin{figure}[h]
    \centering
    \subfigure(a){\includegraphics[width=.5\columnwidth,clip,trim=240 250 250 250]{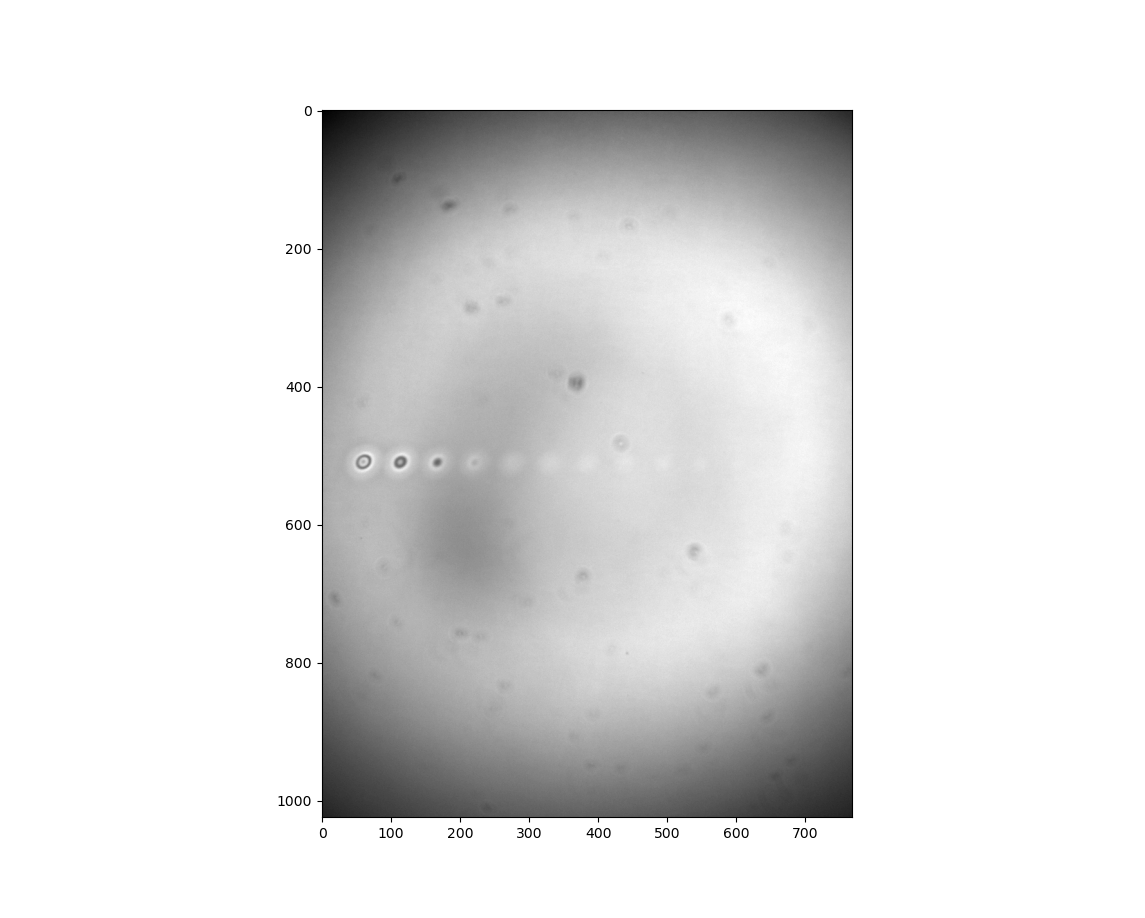}}
    \subfigure(b){\includegraphics[width=.35\columnwidth,clip,trim=10 10 10 10]{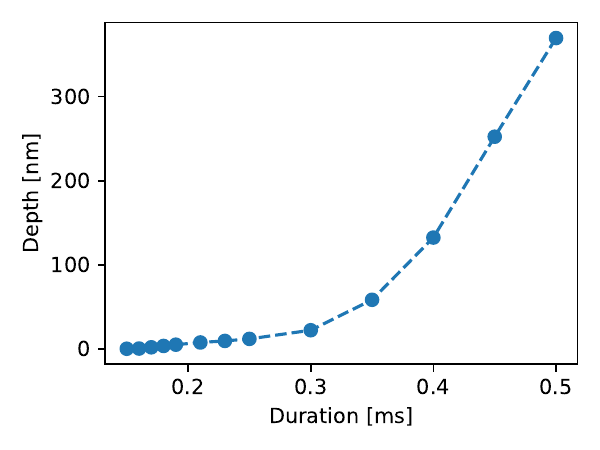}}
    \caption{
      (a) An example depth calibration showing milling spots from 13 pulses at a set power with durations varying between $150\us$ and $500\us$. (b) The calibration curve generated from the milled array. 
    }
    \label{fig:depth_cal}
\end{figure}

Before milling, the relationship between $\rm CO_2$ laser pulse 
parameters and the parameters which describe the resulting surface
indentation are measured.  This calibration is achieved by applying a 
series of $\rm CO_2$ laser pulses to a sacrificial target made of the 
same or similar material to the one used during the iterative refinement milling, see Fig.~\ref{fig:depth_cal}.
The pulses are spatially separated so that the 
indentation formed by each pulse is clearly resolved.  The
pulses are applied with varying duration but the same laser power.
Each indentation is least-squares fit to a Gaussian, and the widths and 
height of the Gaussian are recorded.  The iterative refinement 
algorithm interpolates the calibration data when calculating the set 
of pulse parameters which will minimize the mill error.

Initial error in calibration as well as additional variability due to uncontrolled thermal and mechanical effects relevant at the time scale of multiple milling iterations, change the material removal of a laser pulse. As this scales with the amount of material removed, these errors will approach the level of the constant errors as the measured surface converges to the target surface. 

Experimentation with calibration has revealed that the
volume of the material removed during a laser pulse depends on the
gradient of the surface, as well as previous nearby ablations.  This indicates a complex material response that precludes an effective single calibration and invites a lower than unity proportional gain to increase convergence stability. 

\subsubsection{Pulse Duration Noise}

The $\rm CO_2$ laser is pulsed using an \ac{AOM}.
Although the timing noise does not depend on 
the duration of the laser pulse, the milling error caused by timing 
noise is proportional to the duration of the pulse.  Due to 
the non-linear relationship between the pulse duration and volume of 
material removed (see Fig.~\ref{fig:depth_cal}), this error asymptotically approaches zero as the pulse 
duration approaches zero.

\subsubsection{Laser Power Noise}

The volume of material removed by a laser pulse strongly and non-linearly depends on 
the laser power.  The milling apparatus uses a $5 \W$ $\rm CO_2$
laser (Access Lasers, model L5SLT-AOM), which produces
approximately $1 \W$ at the point of milling.  In our current setup,
iterative refinement with $10-15$ iterations at a grid spacing of $3\um$ over a $50\um$ radius
typically takes $1-2$ hours, while the $\rm CO_2$ 
laser power drifts on a time-scale of several minutes.

To monitor laser power, a pick-off, located down stream of the  
\ac{AOM}, directs laser light onto a photodiode (Thorlabs, PDAVJ10).  
Software PID feeds back onto the \ac{AOM} to stabilize the laser 
power. This is done both when the mill is idle, using a shutter located before the sample, as well as during milling. 
Uncontrolled thermal drifts in our photodiode are currently limiting the efficacy of PID stabilization.

\subsection{Constant Errors}

Constant errors do not scale with the amount of removed material and limit the final achievable
surface error. They include: lower limit calibration error, laser pulse positioning 
error, optical profilometry measurement error, and 
the basis set limitations of a finite pulse grid.  

\subsubsection{Lower Limit Calibration Error}

While the calibration error, as reported by the least-squares fit, is
proportional to the volume of the Gaussian, this error is
still present in, and limited by, the smallest measurable Gaussian indents we can achieve in calibration. This lower limit to calibration sets the minimum removable volume for the milling operation, limiting the minimum milling error.

\subsubsection{Laser Pulse Position Error}

Our system uses a 6-axis nanopostioner hexapod (SmarAct P-SLC-2) to
manipulate the position of the sample relative to the $\rm CO_2$ focal spot
while the laser optics remain static.
The hexapod achieves $\pm15~\nm$ 
of position accuracy over $1~\mm$ of travel, as well as  $\pm10~\um$ 
rotational accuracy over the full rotational range.  The hexapod can  
travel $79~\mm$ along a single axis, which enables the target to move 
between $\rm CO_2$  focus and an optical profilometer while 
maintaining sub-micrometer position accuracy. Choosing an appropriate stage acceleration is important to realize precise movements and reduce mechanical excitation induced positional errors.

The iterative refinement milling method requires precise targeting of 
specific surface locations as measured by the optical 
profilometer.  Small positioning errors result in milling errors 
which do not decrease with milling iteration, while large offset 
errors result in the creation of a positive feedback loop 
leading to the destruction of the sample.

The offset between the location of the sample in the optical profilometer 
and the $\rm CO_2$ focus is measured and updated through a process 
which we perform on each sample before milling.  The sample 
surface is measured before and after firing a laser pulse targeting the 
center pixel in the optical profilometer imaging system.  The 
resulting indentation is least-squares fit to a Gaussian, and corrections 
to the offset are extracted from the position fit parameters.  As the 
iterative refinement milling method does not require a flat initial sample, 
the offset calibration is performed in situ on the mill sample.  This 
calibration is often repeated a few times before beginning the milling process
as this has been found to reduce the offset error. 
The offset between the optical profilometer and the laser focus does 
not significantly drift during sample loading or between milling 
runs. 

To avoid random position error caused by vibrations, the mill is 
isolated from the environment using an actively controlled vibration 
stabilization platform
along with an acoustic dampened and thermally stabilized enclosure.

\subsubsection{Profilometry Measurement Error}
\label{sec:profileometry_error}

\Ac{PSI} extracts profilometry data from interferograms by a
Mirau imaging system.  Phase shifting is achieved by translating 
the target, mounted directly to the hexapod stage, while a phase unwrapping algorithm~\cite{Malacara2007} is used to calculate a surface 
profile. This is used to construct error maps to inform the 
parameters for the next milling iteration or for general calibration
purposes.
We primarily used the Hariharan algorithm~\cite{Malacara2007} due to its simplicity and reasonable results. 

Our custom built \ac{PSI} system's resolution is limited by our camera resolution, 
interferometry wavelength, and choice of phase unwrapping algorithm. This effectively limits the minimum measurable \ac{RMS} 
error to $~200~\picm$, however, we observe single measurement errors closer to 5~\nm, likely due to camera read noise. To address this, we take the average of 20 profile images to produce the profile used to generate the error maps for iteration, reducing this read noise to near 1~\nm. We additionally are limited in the slope of 
surfaces which we can mill, ultimately by the Nyquist limit, 
constraining the arbitrariness of accessible geometries. As the pixel 
size of our camera is $0.28~\um$, and the single wavelength of our 
interferometry light is $477~\nm$, the steepest slope we could 
potentially resolve is roughly $40^{\circ}$. 
In practice our mill geometries are far shallower than this, and we have not rigorously tested more extreme cases.

\subsubsection{Basis Set Limitations}

Our feedback loop attempts to minimize errors by applying corrective Gaussian
indentations to the measured surface at each milling grid point. The
Gaussian parameters are fit from our depth calibrations (see
Sec.~\ref{sec:calibration}).
Due to the
finite number of pulse dots in our grid, we do not have access to a 
complete basis required to analytically describe the 
target surface as a convolution of Gaussians.
This will determine a limiting surface error value 
for a given geometry. This error is generally on the order of 100~\picm for spherical surfaces, though will depend greatly on the grid density and the target surface.

\subsection{Feedback Stability}\label{sec:stability}

During each refinement iteration, we remove a volume of material that 
is proportional to the error between the measured surface and the 
target surface. This proportional feedback gain is set to a value $<1$ to 
prevent instability in the the feedback loop.

\begin{figure}[h]
    \centering
    \includegraphics[width=.9\columnwidth]{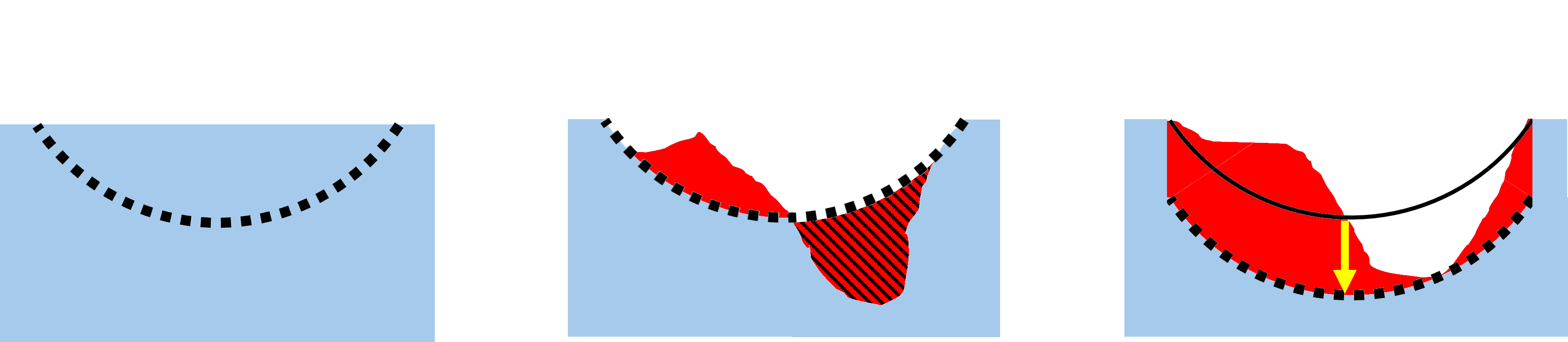}
    \caption{When negative errors (hashed red region) are achieved by milling, our algorithm will shift the target fit down in space, as our system cannot add material to fill in negative errors. This shift enables a convergent approach to the target surface.}
    \label{fig:neg_error}
\end{figure}

As we can only remove material, we don’t see oscillations when the proportional feedback
coefficient is too large.  To correct errors that arise when too much material is removed,
we adjust the height of the target surface so that all errors can be addressed by the
removal of material , see Fig.~\ref{fig:neg_error}.  Therefore a modest and localized positive error, can
necessitate that a large volume of material must be removed across the entire surface of the
target.	Consequently, a positive feedback can arise when the proportional feedback gain is
set too high.

\section{Results}
\label{sec:results}
We demonstrated milling individual spherical, multiple spherical, and conical axicon surfaces using our iterative refinement milling method. We show both phase profile images from our \ac{PSI} as well as calculated residual error maps from representative mills for each design. We additionally demonstrate the ability to mill on various substrate materials, including fused silica slides, \ac{GRIN} fibers, and coreless fibers. 

\begin{figure}[h]
    \centering
    \includegraphics[width=\columnwidth]{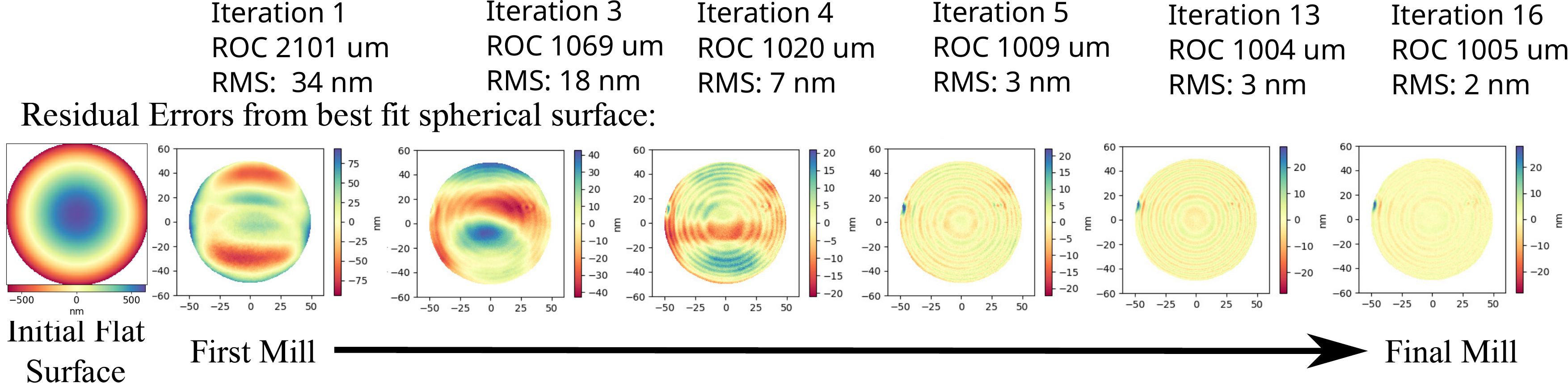}
    \caption{
      Demonstration of the iterative milling of a 1~\mm radius of curvature spherical surface on a
      fiber tip.  The fiber was initially polished.  The plots show the
      measured residual errors of the surface as compared to the exact target
      spherical surface.
    }
    \label{fig:fiber-iteration}
\end{figure}

\subsection{Spherical Mirrors}

Our iterative method has achieved target surface \ac{RMS} of 2~\nm on 100~\um
diameter mirrors with 1~\mm \ac{ROC} on the tips of coreless fiber. This compares favorably with the 12~\nm mean deviations reported by~\cite{Ott2016} using the dot-milling method for mirrors of similar parameters on a similar substrate. Fig.~\ref{fig:fiber-iteration} shows the measured surface profile while applying iterative refinement milling to the tip of an optical fiber across 20 iterations. 
\ac{RMS} values are calculated from our profilometry algorithms 
applied to the entire mirror surface. For high finesse cavities, the 
beam spot size at the mirror will be significantly smaller than
the mirror diameter, indicating that surface error will be most 
impactful near the center of the mirror. 
When limited to a radius of $16~\um$, roughly equivalent to the calculated beam spot size at the mirror for a confocal setup, our iterative milling process 
produces $100~\um$ diameter mirrors with \ac{RMS} surface deviation of 0.77~\nm. Identifying this small \ac{RMS} with surface roughness corresponds roughly to a predicted finesse of $\mathcal{F} \approx 20000$ and an atom cooperativity of
$\eta = 19.6$ in a confocal setup for our mirrors. We have not measured actual finesses for our mirrors. Results from finesse measurements on mirrors produced via dot-milling by~\cite{Ott2016} indicate surface roughness on the order of 0.3~\nm, as determined through back-calculations of cavity losses from a reported finesse of $\mathcal{F} \approx 50000$.   

\subsubsection{Various Materials}

\begin{figure}[ht]
    \centering
    \subfigure(a){\includegraphics[width=.24\columnwidth,clip,trim=190 40 860 300]{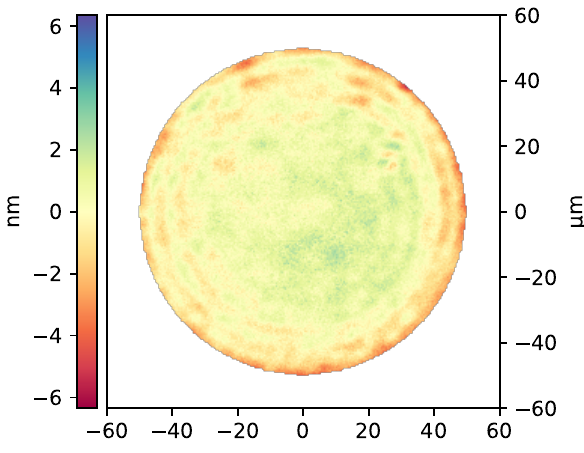}}
    \subfigure(b){\includegraphics[width=.25\columnwidth,clip,trim=180 40 860 300]{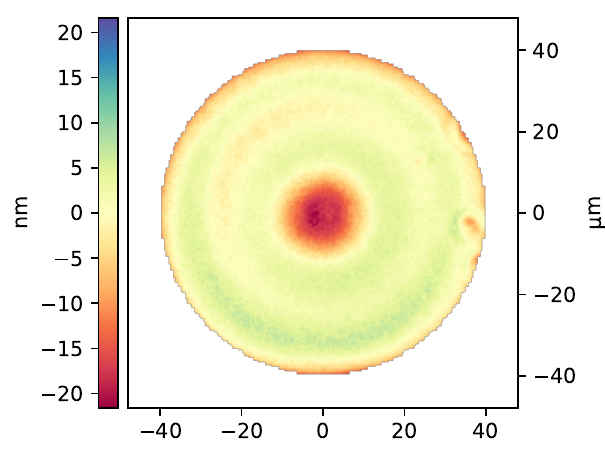}}
    \subfigure(c){\includegraphics[width=.25\columnwidth]{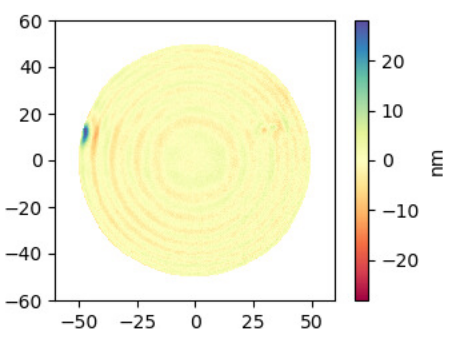}}
    \caption{
      (a) An example mill of a 1~\mm \ac{ROC} spherical surface on a glass slide, 1~\nm \ac{RMS} achieved after 25 iterations. (b) An example mill of a 1~\mm \ac{ROC} spherical surface on the tip of a GRIN fiber, 4~\nm \ac{RMS} achieved after 25 iterations. The material was more readily sculpted than coreless fiber or the glass slide, but our iterative method was able to accommodate and still produce a low roughness surface. (c) An example mill of a 1~\mm \ac{ROC} spherical surface on a coreless fiber tip, 2~\nm \ac{RMS} achieved after 16 iterations. 
    }
    \label{fig:coreless_plus_GRIN_data}
\end{figure}

We have successfully milled cavity mirrors into fused silica slides, coreless fibers (Thorlabs FG125LA), and \ac{GRIN} fibers (Thorlabs GIF50C), see Fig.~\ref{fig:coreless_plus_GRIN_data}. Each of these targets exhibit 
different material responses to the milling process which are able to 
be accommodated for by our iterative method.

Milling \ac{GRIN} fiber tips was explored due to mode-matching concerns in
optical cavity applications. An appropriate length \ac{GRIN} fiber spliced onto a single
mode fiber acts as a lens, enabling tuning of the coupling into the fundamental mode of the
cavity. This increases coupling efficiency by an order of magnitude. 

We additionally attempted to mill on borosilicate glass slides. We found that the material response was drastically different from other targeted materials, demonstrating a limit to the immediate adaptability of our process to material properties. Initial mill attempts elicited material expansion without appreciable removal. Additional calibration and tuning of parameters may still enable successful milling of borosilicate. 

Attempts to mill \ac{MM} fibers highlighted additional calibration complexities. The core and cladding regions of the fiber milled at significantly different rates, precluding effective milling accross the boundary without further calibrations. This result is in agreement with dot-milling results on fibers with core in~\cite{Ott2016}.  

\subsection{Demonstration of Arbitrary Surfaces}

The signature strength of our iterative refinement milling method is the 
flexibility to quickly and efficiently produce optical surfaces with 
arbitrary geometries. We have modeled 
and fabricated several technically relevant examples to demonstrate 
this capability: narrow separation spherical mirrors in various 
configurations and axicon mirrors. Multiple spherical mirrors may enable 
evanescent coupled \acp{FFPC} for use in optical many-body \ac{cQED} 
applications~\cite{Kollar2019} whilst axicon mirrors can enable Bessel beam production for use in
optical tweezing, optical vortex generation, etc.~\cite{Khonina2020}.

We milled closely spaced spherical mirrors with the goal of enabling 
coupled cavity arrays. Evanescent coupled arrays such as these can 
present many-body effects analogous to topological phases of matter, 
useful in exploring protected quantum phases for applications 
including quantum sensing and quantum computing~\cite{Kollar2019}. The example geometry 
of a kagome lattice (tri-hexagonal tiling) was chosen as it presents topologically 
non-trivial band structures, both electronic as well as photonic.

\begin{figure}[h]
    \centering
    \subfigure(a){\includegraphics[height=.15\columnwidth,trim=575 250 542 305,clip]{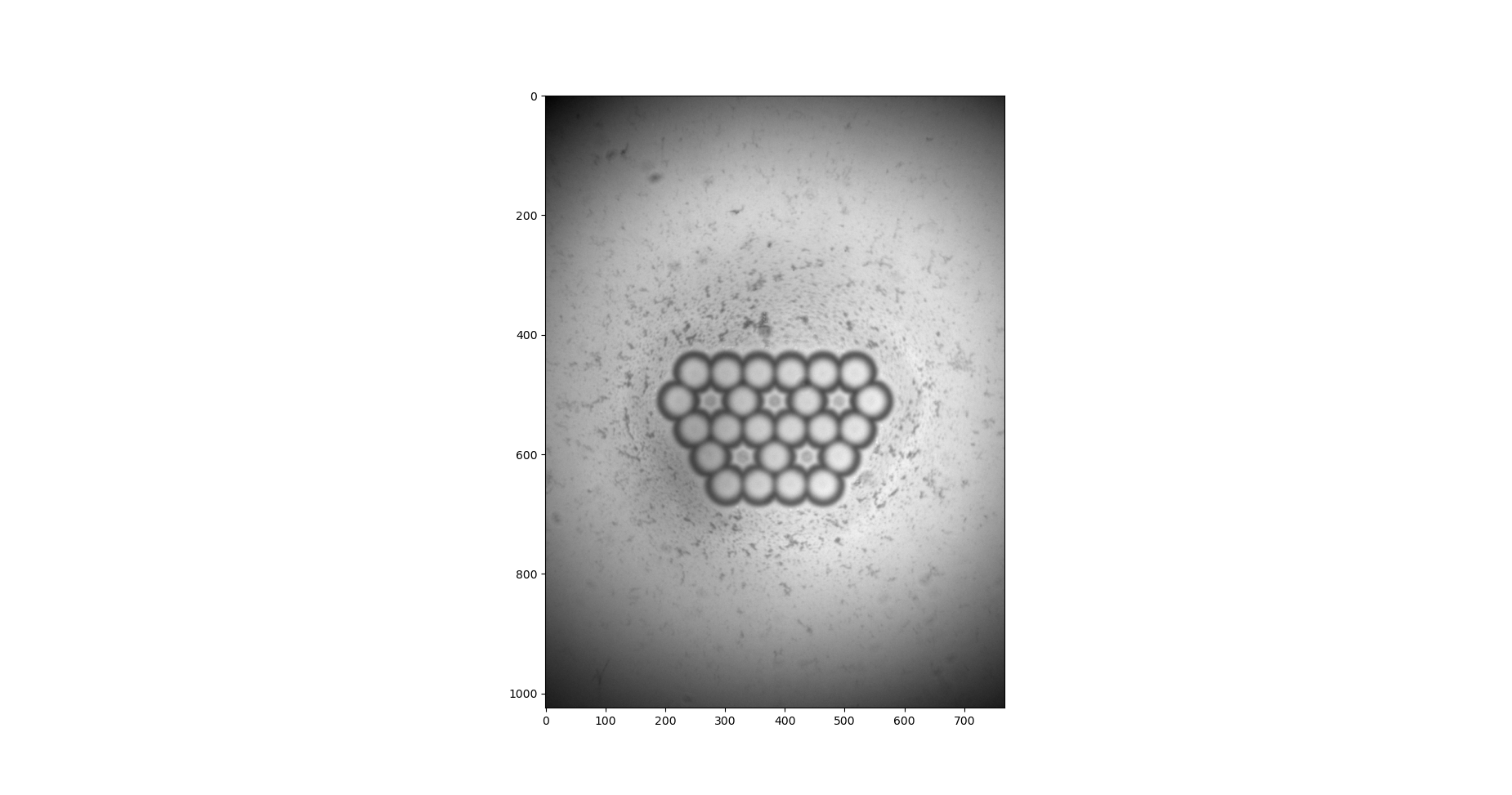}}
    \subfigure(b){\includegraphics[width=.24\columnwidth,clip,trim=70 30 420 230]{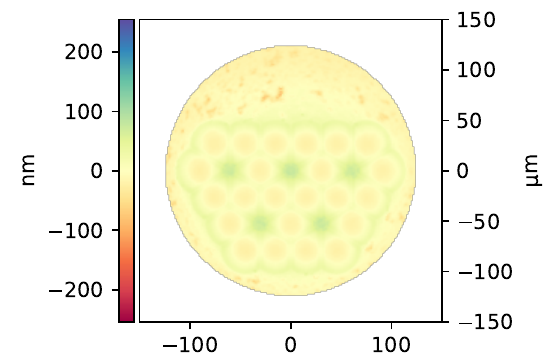}}
    \subfigure(c){\includegraphics[width=.24\columnwidth,clip,trim=70 30 420 230]{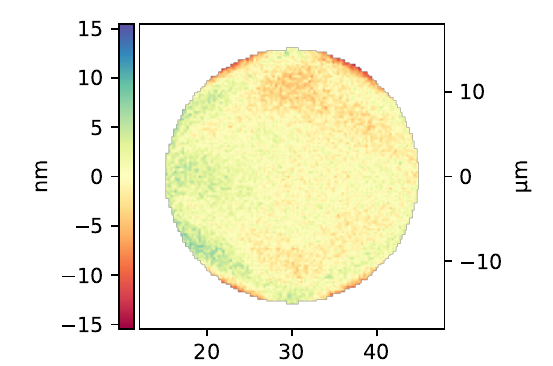}}
    \caption{
      (a) An example mill of a kagome lattice segment of 23 $1.5\mm$ \ac{ROC}
      $25\um$ radius spherical surfaces on a fused silica slide. (b) The residual error of the resultant surface to the target surface, with $19\nm$ total
      surface \ac{RMS} achieved after 30 iterations. (c) The residual error for a single component sphere, with $3\nm$ rms in the central $15\um$ of each sphere. 
    }
    \label{fig:kagome_data}
\end{figure}

In our initial testing, we produced a kagome lattice segment with 23 
spherical mirrors, each with a $1.5~\mm$ \ac{ROC} and a $25~\um$ 
radius, spaced with a $30~\um$ lattice constant. The resultant test
structure was achieved in 30 iterations with a total \ac{RMS}
surface error of $19~\nm$ as shown in Fig.~\ref{fig:kagome_data}. The \ac{RMS} deviation of an individual mirror was 3~\nm. 

\begin{figure}[ht]
    \centering
    \subfigure(a){\includegraphics[width=.17\columnwidth,clip,trim=140 87 140 80]{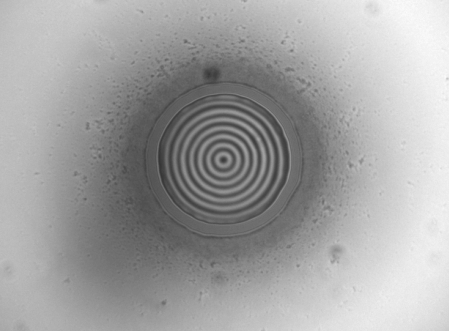}}
    \subfigure(b){\includegraphics[width=.23\columnwidth]{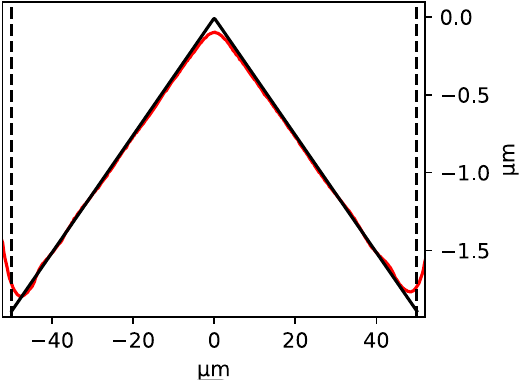}}
    \subfigure(c){\includegraphics[width=.23\columnwidth,clip,trim=70 30 420 240]{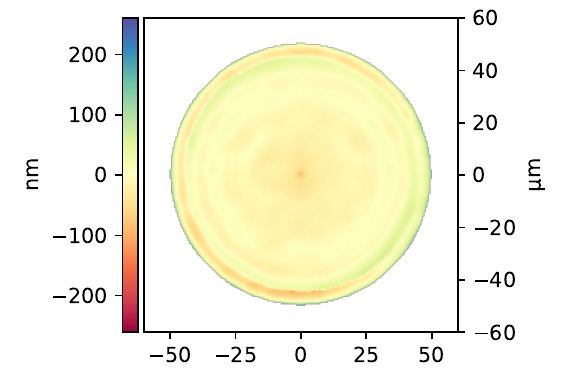}}
    \subfigure(d){\includegraphics[width=.23\columnwidth,clip,trim=70 30 420 240]{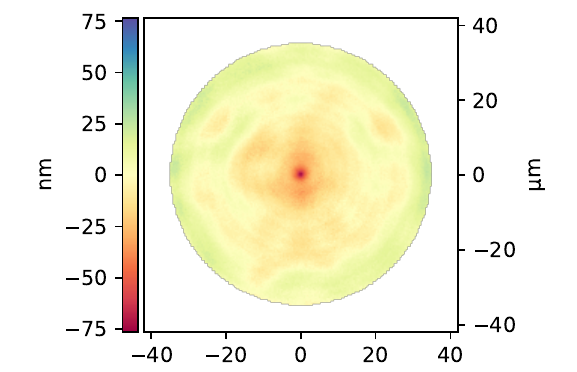}}
    \caption{
      (a) An $87.5^{\circ}$ half-angle conical axicon milled into a fused silica slide, $35\nm$ total surface \ac{RMS} achieved after 20 iterations.
      (b) Cross section of the axicon (red) comparing to target surface (black).
      (c) Residual error of the resultant surface to the target surface showing $35\nm$ \ac{RMS} error.
      (d) Residual error when compared to a $35\um$ reduced-radius target surface showing $10\nm$ \ac{RMS} error; the error is dominated by the rounded tip.
    }
    \label{fig:axicon_data_sim}
\end{figure}

We additionally produced a 100~\um diameter conical axicon with an $87.5^{\circ}$ 
half angle and $35~\nm$ \ac{RMS} error in 20 iterations. A significant portion of this 
surface error was attributable to limitations on the tip sharpness, as shown in Fig.~\ref{fig:axicon_data_sim}.  The
approximately $7~\um$ radius rounded tip effectively limits the
minimum input beam radius such that an appreciable portion of the 
beam will be translated into a Bessel beam. Slightly reducing the fit radius of the surface leads to an \ac{RMS} of 10~\nm, within an order of magnitude of focused ion beam milling fabrication quality~\cite{Cabrini:2006}. 

Our current laser mill setup sets limitations for feature sizes, see 
Sec.~\ref{sec:profileometry_error}, though this only limits the half-angle of the
axicon cone to $40^{\circ}$. We have not yet tested the maximum 
achievable angle for this design.

\section{Conclusions}\label{sec:conclusion}

We have demonstrated the capability to mill high quality, low roughness surfaces 
with arbitrary geometries on various materials, both on fiber 
tips as well as optical flats. We furthermore have shown the ability 
to do so using an agile, iterative method to dramatically reduce the 
time and effort needed to realize a desired surface. 
Our spherical surfaces were demonstrated to have lower \ac{RMS} surface deviation than the dot-milling method. 
Although we have not measured finesses to back-calculate roughness, our current worst-case surface roughness measurements still afford sufficient predicted finesse for various applications. These results represent an 
improvement in manufacturing efficiency and quality over existing technologies, 
and a capability to generate manifold tools for cold atom 
manipulation and \ac{cQED} applications.

This work was funded by the Air Force Office of Scientific Research 
under lab task 22RVCOR017.

We would like to thank Professor Chandra Raman and Jacob
Williamson for their very useful discussions which were helpful in
ferreting out technical challenges. We additionally would like to thank Logan Mamanakis for his assistance upgrading the apparatus.  

\section*{Disclaimer}
The views expressed are those of the authors and do not necessarily 
reflect the official policy or position of the Department of the Air 
Force, the Department of the Defense, or the U.S. Government. The authors declare no conflicts of interest. Data underlying the results presented in this paper are not publicly available at this time but may be obtained from the authors upon reasonable request. 

\clearpage

\bibliography{refinement}

\clearpage

\end{document}